\documentstyle[12pt,aasms4]{article}

\begin{document}

\noindent
{\it Dissertation Summary}

\begin{center}

\title{\large \bf On the Formation and Evolution of Stellar Bars in Galaxies\footnote{See
http://www.astro.iag.usp.br/$\sim$dimitri/phdthesis/phdthesis.html where this work, originally written in
Portuguese, is available.} }

\end{center}

\author{ Dimitri Alexei Gadotti }

\affil{ Astronomy Department - University of S\~ao Paulo, Rua do Mat\~ao, 1226, S\~ao Paulo/SP, Brasil,
05508-900 }

\begingroup

\parindent=1cm

\begin{center}

Electronic mail: dimitri@astro.iag.usp.br

Thesis work conducted at: Astronomy Department - University of S\~ao Paulo

Ph.D. Thesis directed by: Ronaldo Eust\'aquio de Souza ;  ~Ph.D. Degree awarded: 2003 December

{\it Received \underline{\hskip 5cm}}

\end{center}

\endgroup


We have done a detailed study on the structural and kinematical properties of lenticular
and early- and late-type spiral galaxies with bars, aiming to explore the formation and evolution processes
of stellar bars in galaxies, and their implications on the global formation and evolution of galaxies.

Firstly, using high signal-to-noise spectra obtained along the major and minor axes of the bars in a sample
of 14 face-on galaxies, we have determined the line of sight stellar velocity
distribution in the bars' vertical axis, in several points, reaching 20'' from the center. This was done with
an algorithm properly developed for this task, which parameterizes the velocity distribution using Gauss-Hermite series
(see, e.g., R. P. van der Marel \& M. Franx 1993, \apj, 407, 525),
allowing for an accurate determination of the kinematical parameters. These spectra were observed with the
1.5 m ESO telescope at La Silla, Chile, and with the 2.3 m Steward Observatory telescope on Kitt Peak, Arizona.
With these data, it was possible to develop a diagnostic tool that allows one to estimate the ages of bars, and distinguish
between recently formed bars and evolved bars. Furthermore, we could evaluate the vertical structure of disks
and bars in galaxies. We were able to separate evolved bars from recently formed ones based on the
assumption that bars form within disks, thus having initially a thin vertical structure, which can be recognized
by low values for the vertical stellar velocity dispersion, $\sigma_z$. As bars age, processes like vertical resonances
(e.g., F. Combes \& R. H. Sanders 1981, \aap, 96, 164) and the hose instability (e.g., D. Merritt \&
J. A. Sellwood 1994, \apj, 425, 551) contribute to make $\sigma_z$ higher, i.e., turning them vertically thick
and possibly originating the characteristic boxy/peanut morphology (see Figure 1).

Secondly, through realistic $N$-body simulations with {\sc nemo} (P. J. Teuben 1995, ASP Conf. Ser., 77, 398),
which represent a $2-3$ Gyr evolution of isolated galaxies with bulge, disk and a rigid dark matter
halo, we have studied the necessary conditions to the formation of bars in galaxies with different morphological
types (i.e., different bulge/disk ratios). In this way, we could also check the time scales involved in the processes
which give bars an important vertical structure. These simulations show that the current scenario for bar formation
(i.e., via the dynamical bar mode instability that originates globally in kinematically cold disks)
is not able to explain naturally the existence of bars in lenticular
galaxies, which are kinematically hot stellar systems and have prominent bulges. While bars develop
conspicuously in our $N$-body realizations of late-type spirals, the bar mode instability is suppressed in
the realizations of galaxies with important bulges and kinematically hot disks. We thus suggest a new
mechanism for bar formation in galaxies, which is based on the accommodation of the stellar orbits
within a triaxial, or prolate, dark matter halo, which is sufficiently eccentric (D. A. Gadotti \& R. E. de Souza
2003, \apjl, 583, 75). As we showed with $N$-body simulations which consist of only a spheroidal within
such a halo, this mechanism rapidly produces perennial bars in hot stellar systems, and may be an
explanation for the existence of barred galaxies in which the disk component is almost absent,
a discovery which is also a product of this work (see below). Furthermore, this mechanism may
account for the serious drawbacks confronted in the current scenario when applied to barred early-type
spirals and lenticulars, as the triaxial halo may help in the onset of the bar mode instability in otherwise stable disks.
We have also verified that a typical timescale for the thickening of bars, which cause the boxy/peanut morphology,
is of the order of 1 Gyr. However, the values for the bar $\sigma_z$ in our simulations after 1 Gyr are too low ($<$ 50 Km/s)
when compared to the ones we observed in evolved bars ($\sim$ 100 Km/s). Thus, we suggest that another process
may be playing a relevant role after the vertical resonances and/or the hose instability to make $\sigma_z$ as
high as we observe in evolved bars. We have theoretically verified that the Spitzer-Schwarzschild mechanism,
originally proposed for the Galaxy (see L. Spitzer \& M. Schwarzschild 1951, \apj, 114, 385),
is quantitatively able to explain these observations if we assume that giant molecular clouds are twice as much
concentrated along the bar as in the remaining of the disk. The timescales involved in this process
should be $\sim 5-10$ Gyr, which is in agreement with the age differences between young and evolved
bars we have estimated with optical colors (see below).

Thirdly, using images obtained in $B$, $V$, $R$, $I$ and $K\!s$ for a sample of 19 galaxies, and images obtained
in $R$ for a sample of 51 galaxies, including elliptical galaxies, we have performed a detailed structural analysis in galaxies
covering the whole Hubble morphological sequence. The optical imaging was done with the 0.6 m telescope at the Pico
dos Dias Observatory, Brazil, and with the 1.5 m Steward Observatory telescope on Mount Bigelow, Arizona.
The infrared imaging was done with the 2.3 m Steward Observatory telescope on Kitt Peak, Arizona.
We were then able to determine the structural parameters which better describe the bulge and
the disk in these galaxies, as well as obtaining residual images which reveal important sub-structures. In order to perform
this analysis we have developed a specific algorithm, and built an atlas of structural analysis in galaxies.
This algorithm, named {\sc budda} (which stands for BUlge/Disk Decomposition Analysis\footnote{See
http://www.astro.iag.usp.br/$\sim$dimitri/budda.html for the code's internet site.}) performs a 2D
bulge/disk decomposition on galaxy images, assuming a S\'ersic bulge and an exponential disk.
This part of the work revealed the existence of barred lenticular galaxies whose disk component is
negligible (NGC 4608, NGC 5701 and possibly NGC 2217), as well as a difference in the optical colors
of young and evolved bars which amounts to $B-V=0.4$ mag. According to our knowledge on the
evolution of stellar populations, this difference may be translated to a difference in age of the order of 10 Gyr.
This result mean that bars may be, at least in some cases, a long standing structure. We have also found
a correlation between the bulge S\'ersic index $n$ and the galaxy $B-I$ color gradient, in the sense that
bulges with a more concentrated distribution of mass (i.e., with higher values for $n$) have flatter color gradients.
As shown by D. A. Gadotti \& S. dos Anjos (2001, \aj, 122, 1298), galaxies with flat color gradients are likely
to be the ones in which the bar secular evolution processes related to bulge building are relevant. Thus, the
correlation found corroborates the scenario in which bulges in lenticular and in spiral galaxies are, at least
partially, formed through the secular evolutionary processes in bars. Moreover, our results showed that
longer bars appear exclusively in galaxies with faint disks, that there is a correlation between the color
of the bar and the central disk intensity (in the sense that redder bars are in galaxies with fainter disks), and
that evolved bars are longer than young bars, in average. Altogether, these results seem to indicate that,
during its evolution, a bar grows stronger by capturing stars from the disk, which becomes fainter, in agreement with
recent numerical and analytical results (E. Athanassoula 2003, \mnras, 341, 1179). Thus, another possibility to
explain the existence of barred galaxies with almost no disk is that their bars may have evolved
so strongly by consuming their disks that there is almost no disk left.

Finally, our structural analysis also showed that around 1/3 of the elliptical galaxies harbor inner disks, which may be
the result of recent mergers, and that visual morphological classification of galaxies is wrong in nearly 1/3 of the
elliptical and lenticular galaxies, being equally easy to misclassify an elliptical as a lenticular and the other way around.

This work was supported by FAPESP grant 99/07492-7.

\figcaption[fg1.eps]{Measured values for $\sigma_z$ along the bar major axis of 4 galaxies in our sample. The
two panels at left are examples of recently formed bars while the ones at right are examples of evolved bars.
In NGC 5383, the high values of $\sigma_z$ in the inner region correspond to its bulge, while the drops at each
side of the center are caused by its inner spiral arms. In NGC 5850, an inner bar identified in the {\sc budda}
residual images is the likely reason for the noticeable drop in the center.}

\newpage

\plotone{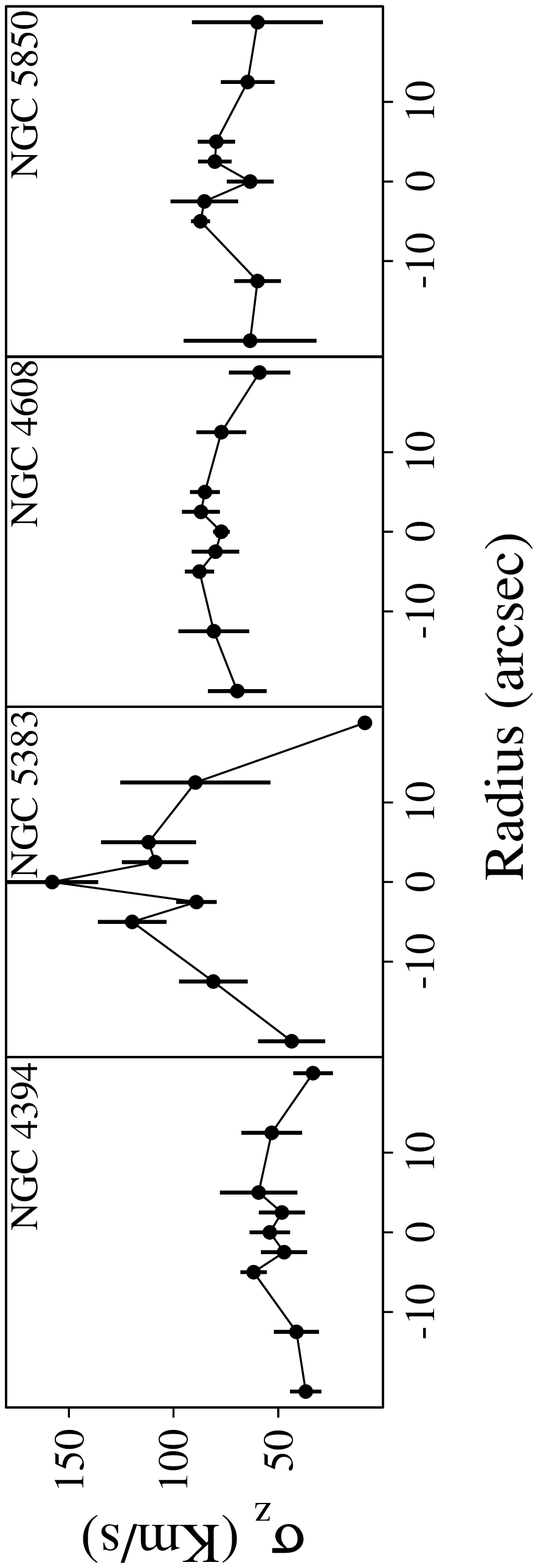}

\end{document}